\documentclass[%
 reprint,
nofootinbib,
 amsmath,amssymb,
 aps,
prx,
]{revtex4-2}

\usepackage{graphicx}
\usepackage{dcolumn}
\usepackage{bm}
\usepackage{xcolor}%
\usepackage{amsmath}%
\usepackage{slashed}%
\usepackage{hyperref}

\begin{document}


\title{Cosmic Shimmering: the Gravitational Wave Signal of Time-Resolved Cosmic Shear Observations}

\author{Giorgio Mentasti}\email{g.mentasti21@imperial.ac.uk}
\author{Carlo~R. Contaldi}%
\affiliation{%
 Blackett Laboratory, Imperial College London, SW7 2AZ, UK
}%

\date{\today}


\begin{abstract}
We introduce a novel approach for detecting gravitational waves through their influence on the shape of resolved astronomical objects. This method, complementary to pulsar timing arrays and astrometric techniques, explores the time-dependent distortions caused by gravitational waves on the shapes of celestial bodies, such as galaxies or any resolved extended object. By developing a formalism based on that adopted in the analysis of weak lensing effects, we derive the response functions for gravitational wave-induced distortions and compute their angular correlation functions. Our results highlight the sensitivity of these distortions to the lowest frequencies of the gravitational wave spectrum and demonstrate how they produce distinct angular correlation signatures, including null and polarisation-sensitive correlations. These findings pave the way for future high-resolution surveys to exploit this novel observable, potentially offering new insights into the stochastic gravitational wave background and cosmological models.
\end{abstract}



\maketitle

\section{Introduction}\label{sec:intro} 
The detection of Gravitational Waves (GWs) by ground-based observatories \cite{PhysRevLett.116.061102} has revolutionised our understanding of the universe, offering a new observational window into the fundamental laws governing its evolution. Ground-based interferometers primarily focus on gravitational waves at frequencies around $f \sim 100$ Hz, produced during the final moments of massive compact binary mergers.
Complementing these observations, the upcoming LISA space mission will explore a different frequency regime, approximately $f \sim 10^{-3}$ Hz, where a wide variety of galactic and extra-galactic sources are expected to exist \cite{Baker:2019nia}.
Additionally, Pulsar Timing Arrays (PTAs) are probing gravitational waves at even lower frequencies, around $f \sim 10^{-8}$ Hz, which has led to recent promising evidence of a gravitational wave background \cite{EPTA:2023fyk, NANOGrav:2023gor,refId0}.

PTAs utilise a network of millisecond pulsars distributed across the Milky Way, with each pulsar acting as an extremely precise cosmic clock. As gravitational waves traverse the galaxy, they induce subtle shifts in the observed pulse arrival times from these pulsars.
By analysing these shifts over time, observatories worldwide collaborate to detect gravitational waves from sources such as supermassive black hole binaries and stochastic gravitational wave backgrounds (SGWBs) in the PTA-sensitive frequency range.

While PTA observations detect the redshift effects caused by GWs, a related method, astrometry, provides an alternative means of detection in the same frequency band \cite{2011PhRvD..83b4024B, Darling2018}.
Astrometric techniques involve precise tracking of apparent angular positions of stars or extra-galactic objects. Gravitational waves passing between an observer and a distant object can cause apparent shifts in the object's position on the celestial sphere.
Upcoming missions, such as Gaia \cite{Gaia}, and surveys like the LSST \cite{LSST:2008ijt}, are ushering in a new era of precision astrometry, enabling the exploration of the gravitational wave universe in ways that supplement traditional techniques. Astrometric observations, particularly those accumulated over year-long timescales, are sensitive to gravitational waves at frequencies similar to those detectable by PTAs.

The rapid advancement of gravitational wave observations, along with the widening range of frequencies under investigation—including emerging efforts at frequencies $f \gtrsim 10$ kHz \cite{Aggarwal:2020olq}—presents exciting opportunities for testing and refining astrophysical and cosmological models. These developments hold promise for addressing key problems in modern cosmology, such as the Hubble tension \cite{LIGOScientific:2017vwq,Branchesi:2023mws,LIGOScientific:2019zcs}, and for probing the nature of dark energy, dark matter, early universe phase transitions, primordial black hole formation, modified gravity models, and inflation \cite{LISACosmologyWorkingGroup:2022jok,LISACosmologyWorkingGroup:2023njw}.

Both interferometric and PTA techniques exploit the effect of gravitational waves—described by irreducible tensorial degrees of freedom—on scalar quantities, such as the Doppler shift or timing offsets they induce. In contrast, the astrometric technique leverages the impact of gravitational waves on a vectorial observable, specifically the angular displacement vector. The projection of tensor metric perturbations in these techniques produces distinct angular responses \cite{PhysRevD.97.124058,OBeirne:2018slh,Inomata:2024kzr,GOLAT2021136381}.

In this {\sl paper}, we introduce an inherently tensorial observable, offering a new class of response functions to gravitational waves that could be exploited in future optical surveys, albeit with much greater resolution and coverage than is possible today. This new observable would provide a novel avenue for detecting low-frequency gravitational waves complementary to PTA and astrometric methods. Specifically, we investigate the time-dependent distortion induced by gravitational waves on the {\sl shape} of resolved astronomical objects. Shape measurements in cosmological observations are a well-established technique.
In cosmic shear studies, the shapes of numerous resolved galaxies are analysed to identify correlated distortions induced by weak lensing \cite{Kaiser:1996tp}, which arises from scalar curvature and potential perturbations along the line of sight between the galaxy and the observer (see \cite{Bartelmann_2001} for a review).  The correlated effect of weak lensing was first detected in the early 2000s \cite{Wittman:2000tc, Hoekstra:2002cj}, and ongoing observational campaigns, such as the Euclid satellite mission \cite{Euclid:2019clj} and the LSST survey with the Vera C. Rubin Observatory \cite{LSST:2008ijt}, now routinely analyse the shapes of billions of galaxies.
These efforts promise to set new precision constraints on dark matter and dark energy models \cite{DES:2021wwk,Euclid:2019clj}. Weak lensing is also routinely observed through its effects on the Cosmic Microwave Background (CMB) anisotropies caused by scalar perturbations along the line of sight to the surface of the last scattering (see, e.g., \cite{Planck:2018lbu}).

Traditional cosmic shear observables are inherently time-independent due to the nature of the underlying weak lensing effect, which produces a constant distortion of background galaxy images. This constancy persists over timescales on the order of the gravitational infall time of the perturbations responsible for the lensing, as well as the homogeneous and isotropic evolution of the cosmological background—both of which far exceed the temporal reach of even long-term (decadal) surveys. In this limit, the perturbations that induce weak lensing possess zero momentum, resulting in an essentially time-independent distortion. Moreover, cosmic shear induced by scalar perturbations results only in magnification and shear of background images, with no induced rotation.

These assumptions are not strictly valid if the weak lensing is induced by vector or tensor perturbations (see, e.g., \cite{Thomas:2009bm}) or if the source of the perturbation has non-zero momentum \cite{Zhu:2014qma}. This work introduces a time-dependent cosmic shear effect induced by gravitational waves.\footnote{To our knowledge, this effect has not yet been proposed as an observational signature of gravitational waves.}
The formalism developed here applies to any small, resolved source in the sky, not only distant galaxies but also galactic objects, which also satisfy the long-distance limit we will be adopting\footnote{The long-distance limit is satisfied if the sources being observed are at distances that are much greater than the gravitational wave wavelength being considered}. The observable proposed here has the distinct advantage that any intrinsic or scalar-induced shear in the background image only induces a mean-field effect and is, therefore, distinguishable in principle from the effect caused by time-dependent gravitational waves. We will show that, despite the inherent observational challenges of this proposed technique, it offers complementary approaches to PTAs and astrometry with distinct angular correlations of the observables as analogues of the Hellings-Down function \cite{HD}.

This {\sl paper} is organised as follows. In section~\ref{sec:distortion}, we briefly review the formalism for describing small distortions of images on the sky, introducing distortion variables analogous to polarisation or shear variables. In section~\ref{sec:formalism}, we derive the distortion induced by gravitational waves and the resulting response functions for each distortion variable. Angular correlation functions for the variables and their spectral analogues are derived in section~\ref{sec:correlators}. We show that convergence and rotation terms lead to the well-known Hellings-Down \cite{HD} function. In contrast, shear variables and their cross-correlations with the scalar observables lead to new correlation functions. We verify our analytical results using a numerical integration scheme. In section~\ref{sec:observations}, we briefly discuss observational prospects. We summarise our findings in section~\ref{sec:discussion}. Detailed derivations of the response and correlation functions are contained in appendices~\ref{app:response_single}-\ref{app:polarised}.

\section{Image distortions}\label{sec:distortion}

We briefly summarise the formalism for considering linearised distortions of background images on the celestial sphere as used to interpret weak lensing observations (see \cite{2020moco.book.....D} for a pedagogical introduction). Consider an image as a flux profile on the sky $I(\vec n)$ where $I$ is an intensity observed in the direction $\hat n$ on the celestial sphere. If the image that we observe is distorted with respect to the background image, we can relate the {\sl observed} and {\sl true} undistorted image as
\begin{align}
    I_{\rm obs}(\hat n) = I_{\rm true}(\hat n + \delta \hat n)\,,
\end{align}
where $\delta \hat n$ is the distortion in the arrival direction of photons induced by the metric perturbations along the line of sight. In the small angle limit and with $|\delta \hat n|\ll 1$, it is convenient to consider the image to lie on the (small) area tangential to the centroid of the observed image. In this case, the mapping between the true image at the source and the observed one can be represented by a linear operator, the distortion matrix 
\begin{align}
    A_{ab}\equiv \frac{\partial (\hat n_a + \delta \hat n_a)}{\partial \hat n^b} = \delta_{ab} + \frac{\partial (\delta \hat n_a)}{\partial \hat n^b}\,,
\end{align} 
with indices $a$, $b$, etc., run over the two orthogonal coordinates that span the plane tangent to the line of sight to the image centroid.\footnote{We use upper and lower indices for consistency, although it is to be understood that contractions are with respect to the Minkowski metric in these indices.} The two-dimensional distortion matrix can be decomposed into irreducible components with respect to rotations as \cite{Stebbins:1996wx} 
\begin{align}\label{eq:psi_ab_decomposed}
A_{ab} = \delta_{ab} + \psi_{ab} = (1-\kappa)\delta_{ab} + \epsilon_{ab}\omega + S_{ab}\,,
\end{align}
where $\epsilon_{ab}$ is the antisymmetric matrix 
\begin{align}
    \epsilon_{ab} = \left(\begin{array}{cc}
         0&  1\\
         -1&0 
\end{array}\right)\,,
\end{align}
and $S_{ab}$ is the irreducible, symmetric, and traceless tensor component, which can be written in terms of two independent degrees of freedom as
\begin{align}\label{eq:Gamma_ab}
    S_{ab} = -\left(\begin{array}{cc}
         \gamma_1&  \gamma_2\\
         \gamma_2& -\gamma_1 
\end{array}\right)\,,
\end{align}
such that
\begin{align}\label{eq:psi}
-\psi_{ab} = \left(\begin{array}{cc}
     \kappa & 0  \\
     0 & \kappa 
\end{array} \right) +
\left(\begin{array}{cc}
     0 & -\omega  \\
     \,\omega & 0 
\end{array} \right)+ 
\left(\begin{array}{cc}
     \gamma_1 & \,\gamma_2  \\
     \gamma_2 & -\gamma_1 
\end{array} \right) \,,
\end{align}
where the contributions have been separated into irreducible scalar, pseudo-scalar, and tensor components. $\kappa$ is the convergence and represents the overall magnification of the image. It contributes to the trace of the distortion matrix. $\omega$ represents rotations of the original image and contributes a transverse, anti-symmetric component. Finally, the $\gamma_1$ and $\gamma_2$ represent the shearing of the image along two sets of perpendicular directions that are offset by 45 degrees with each other and form the transverse, traceless contribution to $\psi_{ab}$. They are defined with respect to the local basis on the tangent plane and are, therefore, inherently coordinate dependent quantities when considered in isolation. 

It is useful to consider how the quantities introduced in \eqref{eq:psi} transform under rotations. This is most conveniently done by considering the spin-weight nature of the distortion components \cite{newman1966note,Goldberg:1966uu}. Spin-weighted quantities can be obtained by contracting irreducible tensors with complex vectors. For traceless, symmetric tensors with two indices, spin $\pm 2$ quantities can be defined by contracting twice with the complex vector constructed using the basis coordinate vectors $\vec m_\pm = {\hat e_\theta \pm i\hat e_\phi}$, 
where, in our case, $\hat e_\theta$ and $\hat e_\phi$ are the orthonormal vectors spanning the tangent plane in a spherical polar coordinate system. The vectors $\vec m_\pm$ transform under right-handed rotations about the line of sight as $\vec m_\pm' = \vec m_\pm\,e^{\mp i\alpha}$ where $\alpha$ is the angle of rotation. Contracting the traceless, symmetric component of \eqref{eq:psi} twice with $\vec m_\pm$ we define two spin-2 quantities
\begin{align}
    {}_{\pm 2}\gamma = \gamma_1 \pm i\gamma_2\,,
\end{align}
that transform as ${}_{\pm 2}\gamma' = {}_{\pm 2}\gamma\,e^{\mp i2 \alpha}$. 

The quantities, $\kappa$, $\omega$, and ${}_{\pm 2}\gamma$ are analogous to Stokes parameters $I$, $Q$, $U$, and $V$ describing the intensity, linear polarisation, and circular polarisation of light respectively \cite{jackson_classical_1999}. These carry the same spin-weights, and much of the formalism described in the following sections is analogous to that used when considering CMB polarisation. As with the case of CMB polarisation, it will be useful to expand ${}_{\pm 2}\gamma(\hat n)$ in spin-2 spherical harmonics \cite{newman1966note,Zaldarriaga:1996xe,Ng:1997ez,Challinor:2002cd,Chon:2003gx,Contaldi:2016koz} when considering the variables and their correlations induced by GWs.

\section{Gravitational wave distortions}\label{sec:formalism}

To calculate the distortion \eqref{eq:distortion} we consider the astrometric deflection of a line of sight along direction $\hat n$ resulting from a gravitational wave $h_{ij}(\hat q)$ with propagation momentum aligned as $\hat k=-\hat q$  \cite{2011PhRvD..83b4024B,PhysRevD.97.124058} 
\begin{align}\label{eq:delta_n}
	\delta \hat n^i (\hat n,\hat q) = \frac{1}{2}\left[ \frac{\hat n^i+\hat q^i}{1+\hat q_l\,\hat n^l}h_{jk}(\hat q)\hat n^j\hat n^k - h^{i}_{\,j}(\hat q)\hat n^j\right]\,.
\end{align}
The astrometric deflection \eqref{eq:delta_n} assumes the signal arriving along the line of sight is sourced at large distances compared to the wavelength of the GWs. In this limit, $h_{ij}$ is understood as evaluated at the observer's position - the ``earth'' term. Source terms arising from the opposite boundary condition can be disregarded in the distant source limit, assuming linearity. As shown in \cite{2011PhRvD..83b4024B}, the expression for the deflection \eqref{eq:delta_n} is also valid at cosmological distances where the comoving radial coordinate differs from the Minkowski radial coordinate.

We can define a general three-dimensional distortion tensor $\widetilde \psi_{ij}$ by taking the covariant derivative of the astrometric deflection with respect to the line-of-sight direction vector. Since we are considering small angular deviations from the line-of-sight at unit radial distances, the derivative reduces to Euclidean.
\begin{align}\label{eq:distortion}
\widetilde \psi_{ij}(\hat n,\hat q)\equiv\frac{\partial \delta \hat n_i}{\partial \hat n^j} &= \frac{1}{2}\left[\delta_{ij}- \frac{\hat n_i+\hat q_i}{1+\hat q_l\,\hat n^l}\hat q_j\right] \frac{h_{rs}\hat n^r\hat n^s}{1+\hat q_l\, \hat n^l}+\nonumber\\
&\frac{\hat n_i+\hat q_i}{1+\hat q_l\, \hat n^l}h_{jr}\hat n^r -\frac{1}{2} h_{ij}\,. 
\end{align}
By introducing spherical polar coordinates centered at the observer's location, we can specify the direction vectors as functions of the angular coordinates of the GW and source
\begin{align}
    \hat q = \cos{\phi}\sin{\theta}\, \hat x + \sin{\phi}\sin{\theta}\, \hat y + \cos{\theta}\, \hat z\,,
\end{align}
and
\begin{align}
    \hat{n} = \cos{\phi_s}\sin{\theta_s}\, \hat x + \sin{\phi_s}\sin{\theta_s}\, \hat y + \cos{\theta_s}\, \hat z\,,
\end{align}
where the subscript $s$ denotes the coordinates of a source along lone-of-sight $\hat n$. 

To recover the tangential distortion matrix \eqref{eq:psi}, we project the three-dimensional tensor using a set of local orthonormal basis vectors in each direction $\hat e_\theta(\hat n),\hat e_\phi(\hat n)$ to obtain a rotated distortion tensor
%
%
\begin{align}\label{eq:psi_ab_notilda}
\psi_{ab}(\hat q,\hat n) = \hat e^i_a(\hat n)\,\widetilde \psi_{ij}(\hat q,\hat n)\,\hat e^j_b(\hat n)\,, 
\end{align}
where $\hat e^i_a$ is the change of basis matrix. Note that since we are considering small distances from the line-of-sight, we can disregard the radial direction by simply letting indices $a$, $b$, etc., run over the two local orthogonal directions $\hat e_\theta$ and $\hat e_\phi$ only and assuming a Euclidean metric. This step fixes the gauge and reduces the distortion to the same form in \eqref{eq:psi}. This determines the calculation of the distortion variables $\kappa$, $\omega$, and ${}_{\pm 2}\gamma$ as a function of the object direction $\hat n$ and GW direction $\hat q$.

\subsection{Distortion response functions}

To calculate the response of distortion variables for a given gravitational wave, we adopt the approach introduced in \cite{allen2024pulsar} where an explicitly diagonal form of the redshift response is obtained by considering the invariance of the response with respect to rotations of the gravitational wave direction $\hat q$. We extend this approach to distortions and show that the diagonal form also exists for the non-scalar quantities ${}_{\pm 2}\gamma$. To do so, we consider the response of the distortion tensor \eqref{eq:distortion} to a left-circularly polarised gravitational wave of unit amplitude, with momentum direction $\hat q$ defined using the local orthogonal basis 
\begin{align}
v^L(\hat q) &= \hat e_\theta(\hat q) + i\hat e_\phi(\hat q)\,,
\end{align}
such that
\begin{align}
    h^L_{ij}(\hat q) &= [v^L\otimes v^L]_{ij}(\hat q)\,,
\end{align}
giving 
\begin{align}\label{eq:distortion_L}
\widetilde \psi_{ij}^L &= \frac{1}{2}\left[\delta_{ij}- \frac{\hat n_i+\hat q_i}{1+\hat q_l\,n^l}q_j\right] \frac{\left[\hat n\cdot v^L(\hat q)\right]^2}{1+\hat q_l\,\hat n^l}+\nonumber\\
&\frac{\hat n_i+\hat q_i}{1+\hat q_l\,\hat n^l}\left[\hat n\cdot v^L(\hat q)\right] v^L_j(\hat q) -\frac{1}{2} h^L_{ij} (\hat q)\,.
\end{align}
To calculate the quantities $\kappa$, $\omega$, $\gamma_1$, $\gamma_2$, we chose local basis vectors defined with respect to tangents to the geodesic connecting the GW direction and the source direction
\begin{align}\label{eq:ephi}
\hat e_\phi(\hat q,\hat n)=\frac{\hat n \times \hat q}{\sqrt{1- (\hat n\cdot\hat q)^2}}\,,
\end{align}
and
\begin{align}\label{eq:etheta}
\hat e_\theta(\hat q,\hat n)=\frac{(\hat n \times \hat q) \times \hat q}{\sqrt{1- (\hat n\cdot\hat q)^2}}\,,
\end{align}
such that we can identify the components of \eqref{eq:psi_ab_notilda} in the presence of a left-handed GW background using
\begin{align}
\psi_{ab}^L &= \hat e^i_a(\hat q,\hat n)
\,\widetilde \psi^L_{ij}(\hat n, \hat q)\,\hat e^j_a(\hat q,\hat n)\,,\nonumber\\
&\equiv\begin{pmatrix}
-F_{\kappa}(\hat q,\hat n) -F_{\gamma_1}(\hat q,\hat n) & -F_{\gamma_2}(\hat q,\hat n)+F_{\omega}(\hat q,\hat n) \\
-F_{\gamma_2}(\hat q,\hat n)-F_{\omega}(\hat q,\hat n) & -F_{\kappa}(\hat q,\hat n) +F_{\gamma_1}(\hat q,\hat n)
\end{pmatrix}\,.
\end{align}
Following \cite{allen2024pulsar}, one can show that the distortion variables calculated in a frame where the gravitational wave is travelling in the negative $z$-direction where $\theta=0$ and the phase angle can be chosen arbitrarily such that 
\begin{align}\label{eq:zaligned}
v^L(\hat z) = e^{-i\phi}(\hat x + i \hat y)\,,
\end{align}
can be related to a rotated frame where the gravitational wave is propagating along a general direction $\hat q$ by a simple set of transformations
\begin{align}\label{eq:transformations}
F_{\kappa}(\hat z,\hat n)&=F_{\kappa}(\hat q,\hat n')\,,\nonumber\\
F_{\omega}(\hat z,\hat n)&=F_{\omega}(\hat q,\hat n')\,,\\
F_{{}_{\pm 2}\gamma}(\hat z,\hat n)&=F_{{}_{\pm 2}\gamma}(\hat q,\hat n')e^{\mp 2i\phi}\,,\nonumber
\end{align}
for a particular choice of rotation $R(\phi,\theta,-\phi)$ that gives\footnote{We use the ZYZ convention for rotations $R(\alpha,\beta,\gamma)$ with respect to Euler angles $\alpha$, $\beta$, and $\gamma$ \cite{1988qtam.book.....V}} $\hat q=R\cdot\hat z$ and $\hat n=R^{-1}\cdot\hat n'$. The scalar variables $\kappa$ and $\omega$ are invariant under this choice of rotation, whereas the spin-2 nature of the ${}_{\pm 2}\gamma$ variables adds a phase term to the transformation.

The advantage of the relations \eqref{eq:transformations} is that the response functions can be calculated in a particular frame where the expressions are simplified and lead to diagonal expansions in the harmonic domain.  The choice of frame where $\hat q=\hat z$ leads to 
\begin{align}\label{eq:vars}
F_{\kappa}(\hat z,\hat n)&=-\frac{1}{2}e^{-2i(\phi-\phi_s)}(1-\cos\theta_s)\,,\nonumber\\
F_{\omega}(\hat z,\hat n)&=i\,F_{\kappa}(\hat z,\hat n) \,,\\
F_{{}_{-2}\gamma}(\hat z,\hat n)&=e^{-2i(\phi-\phi_s)}\,,\nonumber\\
F_{{}_{+2}\gamma}(\hat z,\hat n)&=0\,.\nonumber
\end{align}
As shown in appendix~\ref{app:response_single} these lead to the following harmonic expansions in the {\sl general} frame where the gravitational wave is aligned with direction $\hat q$ and where the $\hat q$ and the $\hat n$ dependence are factorised
\begin{align}\label{eq:single_source_diagform}
F_{\kappa}(\hat q,\hat n)&=\sum_{\ell=2}^\infty A_\ell^\kappa\,\sum_{m=-\ell}^\ell  {}^{\,}_{+2}Y^{\,}_{\ell m}(\hat q)\,Y_{\ell m}^\star(\hat n)\,,\nonumber\\
F_{\omega}(\hat q,\hat n)&=\sum_{\ell=2}^\infty A_\ell^\omega \,\sum_{m=-\ell}^\ell {}^{\,}_{+2}Y^{\,}_{\ell m}(\hat q)\,Y_{\ell m}^\star(\hat n)\,,\nonumber\\
F_{{}^{\,}_{-2}\gamma}(\hat q,\hat n)&=\sum_{\ell=2}^\infty A_\ell^\gamma \,\sum_{m=-\ell}^\ell {}^{\,}_{+2}Y^{\,}_{\ell m}(\hat q)\,{}^{\,{}}_{-2}Y_{\ell m}^\star(\hat n)\,,\nonumber\\
F_{{}^{\,}_{+2}\gamma}(\hat q,\hat n)&=0\,,
\end{align}
where we have introduced scalar spherical harmonics $Y_{\ell m}$ and spin-2 spherical harmonics ${}_{\pm2}Y_{\ell m}$ functions \cite{newman1966note,Zaldarriaga:1996xe,allen2024pulsar} and the $A_\ell$ are the diagonal spherical harmonic coefficients. The form of the distortion variables in the $z$-aligned frame \eqref{eq:vars} allows us to calculate the coefficients as
\begin{align}\label{eq:Als}
A_\ell^\kappa&=(-1)^\ell\, 4\pi \sqrt{\frac{(\ell-2)!}{(\ell+2)!}}\,,\\
A_\ell^\gamma&=(-1)^\ell \,8\pi\frac{(\ell-1)!}{(\ell+1)!}\,,
\end{align}
and $A_\ell^\omega=i \,A_\ell^\kappa$.

\section{Correlators and generalised HD curves}\label{sec:correlators}
The results \eqref{eq:Als} allow us to compute the correlation between distortion variables at two separate source positions $\hat n$ and $\hat n'$. As shown in \cite{allen2024pulsar}, the correlation induced by an unpolarised background of GWs, integrated over all GW directions, is obtained by taking the real component of the correlations in the responses for the left polarised GW calculated above.

In the case of a stochastic GW background, the pure tensorial metric perturbation at the observer's location $\vec x$ is written in terms of a superposition of plane waves.
\begin{align}
    h_{ij}(t,\vec x) = \sum_P \int_{-\infty}^{\infty} \!\!\!\,df \int d^2\hat q\, h^{\,}_P(\hat q,f)\epsilon_{ij}^P(\hat q) \, e^{i2\pi f(t+\hat q \cdot \vec x/c)}\,,
\end{align}
where $\epsilon_{ij}^P(\hat q)$ are the two polarisation tensors and the indices $i$, $j$, run over three spatial coordinates. 
The amplitude $h_P(\hat q,f)$ of each GW plane wave is promoted to a stochastic variable, whose statistics depend on the GW frequency, angular direction, and polarisation. For simplicity, in this work, we limit our analysis to the case of a stationary Stochastic GW Background (SGWB) with
\begin{align}\label{PSD_signal}
\langle h^{\,}_P(\hat{q},f)\rangle &=0\,,\nonumber\\
\langle h^{\,}_P(\hat{p},f)h_{P'}^{\star}(\hat{q}',f')\rangle&=\delta_{PP'}\delta(f-f')\times\nonumber\\
&\delta^{(2)}(\hat{q}-\hat{q}')\mathcal{H}_P(f,\hat q)\,.
\end{align}
In the case of an isotropic, unpolarised background\footnote{The generalisation of this treatment to the case of an anisotropic background is interesting from an observational point of view and can be done by following a treatment analogous to what has been done in \cite{allen2024pulsar} for the anisotropic search with PTA. The generalisation to a polarised background (and to non-Einsteinian polarisation) is feasible and follows a procedure similar to the approach adopted by \cite{PhysRevD.101.024038} again for PTA.
The generalisation to the case of a non-stationary background is much more challenging in terms of data analysis techniques, but the core of our results does not change, as will be explained later.}, the GW power spectrum becomes
\begin{align}
\mathcal{H}_P(f,\hat q)=H(f)\,.
\end{align}
We adopt this assumption throughout the following. 
Considering the time-dependence of the observed astrometric deflection of \eqref{eq:delta_n} induced by the SGWB for an observer at the origin, we can write the time-dependent observed correlator of the distortion variables. For example, for $\kappa$ we have \cite{allen2024pulsar}
\begin{align}
\langle \kappa(\hat n,t) \kappa(\hat n',t')\rangle&=\Re\left[\int d^2\hat q\,F^{\,}_{\kappa}(\hat q,\hat n)F_{\kappa}^\star(\hat q,\hat n')\right]\,,\nonumber\\
&=\int df H(f)\,e^{2\pi i f(t-t')}\times\nonumber\\
&\Re\left[\sum_{\ell m}(A^\kappa_\ell)^2Y_{\ell m}^\star(\hat n)Y^{\,}_{\ell m}(\hat n')\right]\,,\\
&\equiv \Gamma^{\kappa\kappa}(\hat n,\hat n')\int df H(f)\,e^{2\pi i f(t-t')}\,.\nonumber
\end{align}
where we have set the observer location $\vec x=0$ for convenience.
The time (frequency) and the angular dependence in the correlators factorise. This result is analogous to that for PTA and is typical for other astrometric calculations \cite{Mentasti:2023gmr}. This is not the case in interferometric observations, where the overlap functions depend non-trivially on the observed frequencies and the network geometry.

More explicitly, we introduced the angular correlation pattern
\begin{align}\label{eq:Gamma_kk_unpol}
\Gamma^{\kappa\kappa}(\hat n,\hat n')&=\Re\left[\sum_{\ell m}(A^\kappa_\ell)^2Y_{\ell m}^\star(\hat n)Y_{\ell m}(\hat n')\right]\,,\nonumber\\
&=\sum_{\ell\geq 0}\frac{2\ell+1}{4\pi}C^\kappa_\ell \,d_{00}^\ell(\beta)\,,
\end{align}
where $\cos\beta=\hat n\cdot \hat n'$ and the $d^\ell_{mm'}(\beta)$ are the small Wigner $d$-matrices. We include a full calculation for all the possible correlators of the elements of the shear tensor in appendix~\ref{app:response_single} and summarise the results for the angular patterns here as
\begin{align}
\Gamma^{\kappa\kappa}(\hat n,\hat n')&=\sum_{\ell\geq 0}\frac{2\ell+1}{4\pi}C^\kappa_\ell \,d_{00}^\ell(\beta)\,,\\
\Gamma^{\gamma\gamma}(\hat n,\hat n')&=\sum_{\ell\geq 2} \frac{2\ell+1}{4\pi} \,C^\gamma_\ell d_{22}^\ell(\beta)\,\Re\left[ e^{-2i(\alpha-\xi)}\right]\,,\\
\Gamma^{\kappa\gamma}(\hat n,\hat n')&=\sum_{\ell\geq 2} \frac{2\ell+1}{4\pi}\,C^{\kappa\gamma}_\ell  d_{20}^\ell(\beta)\,\Re\left[e^{-2i\alpha}\right]\,,\\
\Gamma^{\omega\gamma}(\hat n,\hat n')&=\sum_{\ell\geq 2} \frac{2\ell+1}{4\pi}\,C^{\kappa\gamma}_\ell  d_{20}^\ell(\beta)\,\Im\left[e^{-2i\alpha}\right]\,,\\
\Gamma^{\omega\omega}(\hat n,\hat n') &=\Gamma^{\kappa\kappa}(\hat n,\hat n')\,,\\
\Gamma^{\kappa\omega}(\hat n,\hat n')&=\Gamma^{\omega\kappa}(\hat n,\hat n')=0\,,
\end{align}
The variables $\alpha$ and $\xi$ are angles subtended by the local tangent frame in both source directions with the tangent to the geodesic between the directions \cite{Ng:1997ez}. The correlations are defined by diagonal angular spectra
\begin{align}\label{eq:cls}
    C_\ell^\kappa &= (A_\ell^{\kappa})^2\,,\nonumber\\
    C_\ell^\gamma &= (A_\ell^{\gamma})^2\,,\\
    C_\ell^{\kappa\gamma} &= A_\ell^{\kappa}\,A_\ell^{\gamma}\,.\nonumber
\end{align}

The correlators involving the spin-2 distortions depend on the choice of coordinates that define the $\gamma_1$ and $\gamma_2$. The dependence results from the non-trivial rotation properties of spin $s\ne 0$ variables. In practice, when analysing spin-$s$ variables on the sphere, expanding in spin-weighted spherical harmonics is more useful where frame-independent $E$ and $B$ spectra can be easily defined \cite{Zaldarriaga:1996xe}. 
Here, we consider coordinate-space correlation functions, a common approach for PTA and astrometry analysis, where sparse sky sampling can lead to additional difficulties in interpreting harmonic-space analogues.
The coordinate dependence is manifested in the additional phases determined by the angles $\alpha$ and $\xi$ subtended by the tangent to the geodesic between $\hat n$ and $\hat n'$ directions and the local meridian at both points. For visualisation purposes, this dependence can be removed by rotating variables to the local frames with $\alpha=\xi=0$ \cite{Ng:1997ez, Chon:2003gx}. This simplifies the non-vanishing correlators as
\begin{align}
\Gamma^{\kappa\kappa}(\hat n,\hat n')&\equiv \Gamma^{\kappa\kappa}(\beta)=\sum_{\ell\geq 0} \frac{2\ell+1}{4\pi}C^\kappa_\ell \,P^{\,}_\ell (\beta)\,,\\
\Gamma^{\gamma\gamma}(\hat n,\hat n')&\equiv \Gamma^{\gamma\gamma}(\beta)=\sum_{\ell\geq 2} \frac{2\ell+1}{4\pi} \,C^\gamma_\ell d_{22}^\ell(\beta)\,,\\
\Gamma^{\kappa\gamma}(\hat n,\hat n')&\equiv \Gamma^{\kappa\gamma}(\beta)=\sum_{\ell\geq 2} \frac{2\ell+1}{4\pi}\,\,C^{\kappa\gamma}_\ell  d^\ell_{20}(\beta)\,,
\end{align}
where $P_\ell(\beta)$ are the Legendre polynomials. 

In appendix~\ref{app:polarised}, we show that, as expected, more correlators are non-zero for a circularly polarised background where $H_L(f)\ne H_R(f)$. In analogy with (10)-(13) of \cite{Contaldi:2015boa}, circular polarisation would source contribution analogous to ``$TB$'' and ``$EB$'' cross-correlations.  In the distortion case, these are imaginary contributions to $\Gamma^{\kappa\omega}$ and $\Gamma^{\omega\gamma}$ respectively. Therefore, the distortion correlation functions can discriminate chirality in GW backgrounds.

\begin{figure}[t!]
\centerline{\includegraphics[scale=0.35]{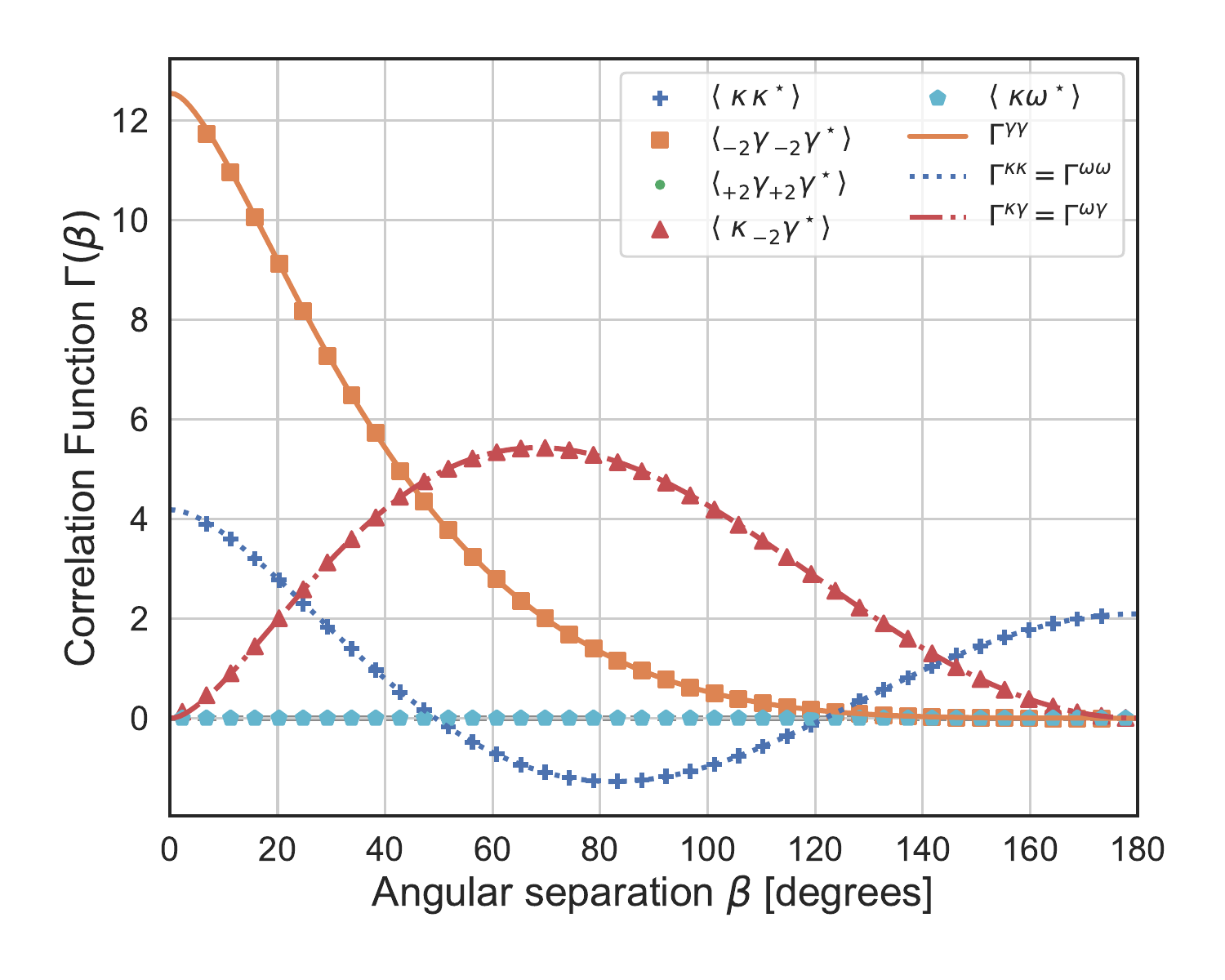}}
\caption{\label{fig:HD}Comparison between the analytical angular correlation functions and their numerical evaluation using a general frame and averaging over random correlation separations. The correlations behave as expected. The $\Gamma^{\kappa\kappa}$ and $\Gamma^{\kappa\omega}$ correlation functions are identical and have the same form as the Hellings-Down angular correlation for redshifting. $\beta$ is the angular distance between the directions being correlated.}
\end{figure}


As a check of the assumptions used in deriving the responses and correlation functions, we evaluate them numerically by approximating the integrals over {\tt HEALPix} maps \cite{Zonca2019}
\begin{align}\label{eq:numerical}
    \int \!\!\! d^2\hat q \,\langle X(\hat q,\hat n)\,Y^\star(\hat q,\hat n')\rangle &\sim  \frac{4\pi}{N_{ij}N_{\rm pix}} \times\nonumber\\
    &\sum_{p,\,i\ne j} \, X(\hat q_p,\hat n_i)\,Y^\star(\hat q_p,\hat n_j)\,,
\end{align}
where $X$ and $Y$ are the distortion variables evaluated using \eqref{eq:psi_ab_notilda} in a general frame where the GW direction is aligned with the radial direction at each of the pixel centres and the basis vectors $\hat e_\theta$ and $\hat e_\phi$ are aligned with the geodesic between the pair of directions being correlated ($\hat n_i$, and $\hat n_j$). This sets angles $\alpha=\xi=0$, which is useful for visualisation.

Since the angular spectra in \eqref{eq:cls} all scale as $\ell^{-m}$ with $m\ge 2$, \eqref{eq:numerical}  converges sufficiently well for relatively low-resolution maps with $N_{\rm side}=16$ and we average over $N_{ij}\sim {\cal O}(10^4)$ pairs of directions. The results are shown in Figure~\ref{fig:HD}, with numerical results matching the analytical curves. The angular correlation for the $\kappa$ and $\omega$ variables have the same form as the Hellings-Down function \cite{HD}. This is not true for correlators involving the shear variables and their cross-correlation with $\kappa$ and $\omega$. 

\section{Observational prospects}\label{sec:observations}

We now estimate the signal-to-noise ratio (SNR) of observables using some simplifying assumptions. A more detailed analysis is beyond the scope of this work and will be left for future investigation.

We start defining a prototype datastream for, e.g., $\kappa$ for the $i$-th object being imaged
\begin{align}
d^i_\kappa(t)=\kappa^i(t)+\kappa^i_0+n^i_\kappa(t)'\,,
\end{align}
where $d^i_\kappa(t)$ is the estimate of the coefficient $\kappa$ at each observation as a function of time $t$. In principle, $d^i_\kappa(t)$ is obtained by considering the components of the second moment of the image fluxes. $\kappa^i(t)$ is the time-dependent effect of the gravitational wave, and $\kappa^i_0$ is the intrinsic effect due to weak lensing, which we assume does not vary over the timescales we are considering.
$n^i_\kappa(t)$ is an effective noise resulting from estimating the $\kappa$ variable from the flux moments. Analogous formulas can be written for $_{\pm 2}\gamma$ and $\omega$. We assume for simplicity that the measurements of the variables $\kappa$, $_{\pm 2}\gamma$ and $\omega$ belonging to the same object are not correlated and have the same order of magnitude in error. However, this may not be true in practice. 

In the noise-dominated case, we assume that the error we introduce when removing the intrinsic bias $\kappa^i_0$ from the data stream is negligible with respect to the noise $n_\kappa^i(t)$.
Furthermore, we assume that the error on the single measurement of $\kappa$ (the same applies to the other coefficients) is white and Gaussian. In the frequency domain, it follows the statistics
\begin{align}
\langle \tilde n^i_\kappa(f)\tilde n^{j\star}_\kappa(f')\rangle =\delta_{ij}\delta(f-f')\sigma_\kappa^2\Delta t\,,
\end{align}
where $\sigma_\kappa$ is the predicted error in each measurement of the coefficients $\kappa$, and $\Delta t$ is the cadence of the experiment.

To estimate this quantity, we can consider the definition of the tensor of the second momenta of the image flux $q_{ij}$ defined as an integral over the flux $I_{\rm obs}(\theta)$ over the cartesian coordinates $\theta=\{x,y\}$ spanning the tangential plane \cite{2020moco.book.....D}
\begin{align}\label{eq:qij_true}
q_{ij}=\frac{1}{I_0\,\Theta^2}\int d^2\theta I_{\rm obs}(\theta)\theta_i\theta_j\,,
\end{align}
where, for simplicity, we assume that the intensity $I_{\rm obs}(\theta)$ is constant over the area covered by the object being observed. We assume the object subtends a solid angle size of $\Theta$.

In practice, the resolution is finite, and we only observe a discretised version of this quantity across $N_{\rm pix}\simeq\Theta^2/\sigma_\theta^2$ effective resolution elements,
\begin{align}
q_{ij}^{\rm obs}=\frac{1}{I_0\,\Theta^2}\sigma_\theta^2 \sum_{n=1}^{N_{\rm pix}}I_{\rm obs}(\theta^n)\theta_i^n\theta_j^n\,,
\end{align}
where $\sigma_\theta$ is the angular size of the pixels representing the angular resolution of the observations.
The error in the measurement of the distortion variables is of the order of
\begin{align}
\sigma_\kappa&=\frac{\left|q_{ii}^{\rm obs}-q_{ii}\right|}{q_{ii}}\,,\nonumber\\
&\sim\frac{1}{I_0\,\Theta^2}\sum_n\frac{\sigma_\theta^4}{24}\frac{6}{\Theta^2}\nabla^2 \left[I_{\rm obs}(\theta^n)\theta_i^n\theta_i^n\right]\,,\nonumber\\
&\sim N_{\rm pix}\frac{\sigma_\theta^4}{\Theta^4}\sim \frac{\sigma_\theta^2}{\Theta^2}\,,
\end{align}
where $q_{ii}\simeq \frac{\Theta^2}{6}$ from the explicit integration of \eqref{eq:qij_true}.

\begin{figure}[t!]
\centerline{\includegraphics[scale=0.35]{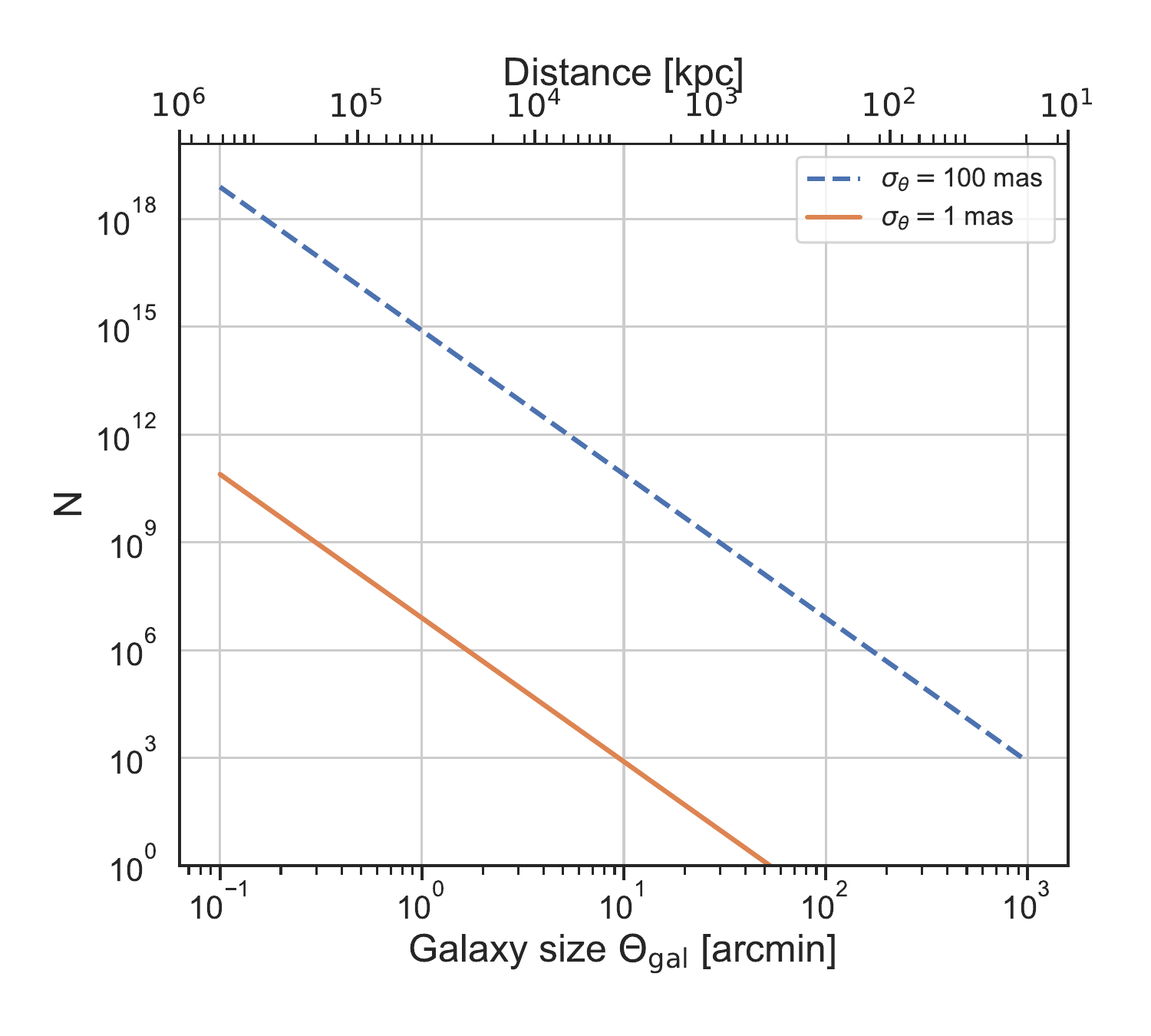}}
\caption{\label{fig:SNR}Number of resolved objects $N$ necessary to produce an $\text{SNR}_{\rm tot}=1$ in \eqref{eq:SNR}. The physical distance to the object is shown on the top axis, assuming they are all galaxies of physical size $D=10$ kpc.}
\end{figure}

The optimally filtered SNR of the signal, considering the measurement of $\kappa$ only in isolation, will be
\begin{align}
\text{SNR}^2_\kappa&\simeq \langle \left|\Gamma^{\kappa\kappa}(\beta)\right|^2\rangle \frac{N^2T}{\sigma_\kappa^4\Delta t^2}\int_{1/T}^{1/\Delta t} df \left(\frac{h_c^2(f)}{16\pi f}\right)^2\,,\nonumber\\
&\simeq 1.5\,\frac{N^2}{\sigma_\kappa^4\Delta t^2}\frac{T}{256\pi^2}\frac{h_{\rm ref}^4}{f_{\rm ref}^{4\gamma}}\int df f^{4\gamma-2}\,,\nonumber\\
&\simeq 1.5\,\frac{N^2}{\sigma_\kappa^4\Delta t^2}\frac{T}{144\pi^4}\frac{h_{\rm ref}^4}{f_{\rm ref}^{4\gamma}}\frac{T^{1-4\gamma}}{1-4\gamma}\,,
\end{align}
where $N$ is the number of objects contained in the survey, $T$ the total time of observation and 
\begin{align}
    h_{c}=h_{\rm ref}\left(\frac{f}{f_{\rm ref}}\right)^{\gamma}\,,
\end{align}
is the spectral form for the characteristic strain typically assumed in PTA analysis. As reference values we adopt $h_{\rm ref}=3\times 10^{-15}$ at $f_{\rm ref}=3\times 10^{-8}$Hz and $\gamma=-2/3$ \cite{NANOGrav:2023hfp,EPTA:2023xxk}. 

Assuming an optimistic $T=15$ year and $\Delta t=3$ days leads to a
the estimate
\begin{align}\label{eq:SNR}
\text{SNR}_\kappa&\simeq 14\left(\frac{0.1 \text{mas}}{\sigma_\kappa}\right)^2\left(\frac{N}{10^9}\right)\,,\nonumber\\
&=4.26\times 10^{-7}\left(\frac{0.1 \text{as}}{\sigma_\theta}\right)^4\left(\frac{\Theta}{1\,\text{arcmin}}\right)^4 \left(\frac{N}{10^9}\right)\,.
\end{align}
The total SNR comes from the cross-correlation of all the possible pairs of $\kappa$, $_{\pm 2}\gamma$ and $\omega$ for all the possible extended objects surveyed. This amounts to 9 possible couples of correlators between each pair of objects and a total SNR of approximately
\begin{align}
\text{SNR}_{\rm tot}\simeq 3 \times \text{SNR}_\kappa\,,
\end{align}
where we assumed for the sake of simplicity that the average angular correlation function and the error in the measurement of the other shear coefficients are of the same order of magnitude as those for $\kappa$.

We show the number of objects of a given angular size required to reach an $\text{SNR}_{\rm tot}=1$ in figure~\ref{fig:SNR}. We can see that a future, high-resolution survey of a limited number of resolved objects could reach sensitivities similar to existing PTA techniques. 

\section{Discussion}\label{sec:discussion}
Using distortion measurements for gravitational wave (GW) detection represents a new avenue for complementing existing methods such as PTAs and astrometric observations. While PTAs capture temporal shifts in pulsar arrival times and astrometry measures angular displacements of celestial objects, our proposed measurements offer a fundamentally different approach by focusing on the tensorial distortions induced by GWs on the shapes of resolved astronomical sources. The novelty of the distortion technique is its ability to disentangle the contributions of GWs from intrinsic or scalar-induced shape distortions. As discussed, weak lensing produces time-independent distortions, whereas GWs induce a time-dependent signal that can be isolated in principle. This makes it possible to distinguish the signal of GWs from other astrophysical effects that cause static or long-term distortions in the shapes of galaxies. The distinct angular correlations of the tensorial distortions could be leveraged to detect the stochastic GW background or specific sources like supermassive black hole binaries.

A significant challenge, however, is the need for high-precision shape measurements and long observational baselines with a high observational cadence to achieve the necessary sensitivity. The error in distortion measurements is currently dominated by the instruments' resolution, with surveys such as Gaia and the Vera Rubin Observatory (LSST) reaching angular resolutions on the order of milli-arcseconds. As our analysis suggests, improving the signal-to-noise ratio (SNR) will require a large sample of low-redshift galaxies with large angular sizes and a high angular resolution. Although not currently practical, future surveys could be designed to optimise sensitivity to this effect. Improvements in optical, astrometric and interferometric techniques will enhance the observational viability of this method.

In summary, while the technique presented here is currently limited by observational limitations, it offers a compelling new tool for gravitational wave astronomy. As technology and data quality improve, distortion-based GW detection or {\sl cosmic shimmering} could significantly enhance our ability to probe the universe, providing insights into the nature of supermassive black holes, the stochastic gravitational wave background, and also cosmological issues such as the Hubble tension, dark energy, and primordial black hole formation. The complementary nature of the observable also lends itself to tests of non-Einsteinian polarisations, anisotropic backgrounds, and cross-correlations with PTA and astrometry measurements, which we will leave for future study.

\begin{acknowledgments}
We thank Alan Heavens for insightful discussions on galaxy shear measurements. G.~M. acknowledges support from the Imperial College London Schr\"odinger Scholarship scheme. C.~R.~C. acknowledges support under a UKRI Consolidated Grant ST/W000989/1.
\end{acknowledgments}

\appendix

\section{Calculation of the distortion response functions}\label{app:response_single}
Following the approach of \cite{allen2024pulsar}, we consider the response to a left-circularly polarised GW of unit amplitude as in \eqref{eq:distortion_L}. It is important to consider how the distortion matrix behaves under rotations of the coordinate systems such that we can relate the response in a preferred frame where the GW is aligned with the $z$-axis \eqref{eq:zaligned} and a general one. As discussed in \cite{allen2024pulsar}, by setting the arbitrary initial phase to $\phi$ and using the rotation $R(\phi,-\theta,-\phi)$ that rotates the $z$-axis to the direction $\hat q$ with angular coordinates $\theta$ and $\phi$, ensures the covariant behaviour of the redshift response function since
\begin{align}
    \hat e_\theta(R\cdot \hat q) + i \hat e_\phi(R\cdot \hat q) = R\cdot\left[\hat e_\theta(\hat q) + i \hat e_\phi(\hat q)\right]\,,
\end{align}
which is not guaranteed in general. Whilst it is not surprising that this observation shows the redshift response function is a scalar under the rotations, it can be extended to vector and tensor objects to ensure they remain covariant under the same rotation. In particular, we can use this same approach to relate $\psi_{ab}(\hat z, \hat n)$ to $\psi_{ab}(\hat q,\hat n')$ and therefore obtain the transformation properties \eqref{eq:transformations} that allow us to use the responses in the preferred frame \eqref{eq:vars} which are analogous to (B6) of reference \cite{allen2024pulsar}. 

Calculating the response functions for the scalar (pseudo-scalar) variables $\kappa$ and $\omega$ proceed similarly. We reproduce the main steps in the calculation here for completeness. The response for a single GW aligned with the general directions $\hat q$ is
\begin{equation}
    F_\kappa(R\cdot \hat z, R\cdot \hat n) = F_\kappa(\hat z, \hat n) = \sum_\ell a_\ell^\kappa \, Y^\star_{\ell,-2}(\hat n)\,,
\end{equation}
and under the redefinition $\hat n \to R^{-1}\cdot \hat n$
\begin{align}
    F_\kappa(\hat q, \hat n) &= \sum_\ell a_\ell^\kappa \, Y^\star_{\ell,-2}(R^{-1}\cdot\hat n)\,,\nonumber\\
    &=\sum_{\ell m} a_\ell^\kappa \, {\cal D}^{\ell\star}_{m,-2}(R^{-1})\,Y^\star_{\ell,-2}(\cdot\hat n)\,,\\
    &=\sum_{\ell m} A_\ell^\kappa \, {\,}_{-2}Y_{\ell m}(\hat q)\,Y^\star_{\ell,-2}(\hat n)\,,\nonumber\\
\end{align}
where the ${\cal D}^\ell_{mm'}$ are the Wigner-D rotation operators and ${\,}_{s}Y_{\ell m}(\hat q)$ are the spin-$s$ spherical harmonics (see e.g. \cite{newman1966note,Zaldarriaga:1996xe,Ng:1997ez}) and
\begin{align}
    a_\ell^\kappa = (-1)^\ell \sqrt{4\pi(2\ell+1)}\sqrt{\frac{(\ell -2)!}{(\ell+2)!}} \, e^{-i2\phi}\,,
\end{align}
and the $A_\ell^\kappa$ is defined in \eqref{eq:Als}. The calculation for $F_\omega$ is similar, and we omit it here. Since the response is identical to the redshift case for $\kappa$ and $\omega$, we obtain the same diagonal form for the general response function. 

In principle, the procedure for $F_\gamma\equiv F_{_{-2}\gamma}$ would proceed in the same way but expand the response over spin-2 spherical harmonics such that
\begin{align}
    F_\gamma(\hat z, \hat n) = \sum_\ell {\,}_{-2}a_\ell^\gamma {\,}_{-2}Y^\star_{\ell, -2}\,.
\end{align}
However, given the more complicated behaviour under the rotation $R^{-1}\cdot \hat n$, it is more convenient to use the spin raising and lower operation to first define rotationally invariant quantities, carry out the calculation, and then use the reciprocal operation to recover the correlation functions. 

Recalling from \eqref{eq:vars} that $F_\gamma(\hat z,\hat n) = e^{-i2(\phi-\phi_s)}$ is the response for the spin-($-2$) variable ${\,}_{-2}\gamma$, we operate twice with the spin {\sl raising} operator acting on the $\hat n$ dependence defined as \cite{newman1966note}
\begin{align}
{\slashed \partial}\, _s f(\theta,\phi) = -\sin^s(\theta)\left[\frac{\partial}{\partial\theta} +  \frac{i}{\sin(\theta)}\frac{\partial}{\partial\phi}\right]\frac{\,_sf(\theta,\phi)}{\sin^{s}(\theta)}\,,
\end{align}
where $sf(\theta,\phi)$ is any spin-$s$ function on the sphere. The double-raising defines an operation that is equivalent to a covariant Laplacian for spin-(-2) functions (see also \cite{Zaldarriaga:1996xe}), which we can use to define a new function
\begin{align}
F_\zeta(\hat z, \hat n)\equiv{\slashed \partial}^2F_\gamma&=\left(-\partial_\mu-\frac{2}{1-\mu^2}\right)^2\left[\left(1-\mu^2\right) F_\gamma\right]\,,\nonumber\\
&=2e^{-2i(\phi-\phi_s)}\frac{1-\cos\theta_s}{1+\cos\theta_s}\,,
\end{align}
where $\mu=\cos \theta_s$\footnote{This simplified operator applies to our case where $\partial_\phi F_\gamma=-2i F_\gamma$.}. 

The advantage of having defined $F_\zeta$ is that it behaves as a scalar under rotations and can be expanded using spin-0 spherical harmonics, as in the calculation above, albeit with a different set of coefficients. The new response function also preserves the convenient property that, in the $z$-aligned frame, the dependence on the phase angle $\phi$ is still trivial, leading to a $\delta_{m,-2}$ in the expansion such that 
\begin{align}
F_\zeta(\hat z, \hat n)&=\sum a_\ell^\zeta \,Y_{\ell -2}^\star(\hat n')\,,
\end{align}
with
\begin{align}
a_\ell^\zeta=2\sqrt{4\pi (2\ell+1)}\frac{(-1)^\ell(\ell-1)(\ell+2)}{\sqrt{\pi(\ell+2)(\ell+1)\ell(\ell-1)}}\,e^{-2i\phi}\,,
\end{align}
with $a^\zeta_{\ell < 2} = 0$, as expected.
The subsequent steps are identical to the previous calculation, and we obtain the diagonal form
\begin{align}
F_{\zeta}(\hat q,\hat n)&=\sum_{\ell m}\, A_\ell^\zeta \,{\,}_{2}Y_{\ell m}(\hat q)Y_{\ell m}^\star(\hat n)\,,\nonumber\\
&=\sum_{\ell m}(-1)^m\, A_\ell^\zeta \,_2Y_{\ell m}(\hat q)\,Y_{\ell -m}(\hat n)\,,
\end{align}
with
\begin{align}
A_\ell^\zeta&=(-1)^\ell \, 8\pi\frac{(\ell-1)(\ell+2)}{\sqrt{(\ell+2)(\ell+1)\ell(\ell-1)}}\,.
\end{align}
To recover the diagonal form for the response function of $F_\gamma$, we define the inverse Laplacian operation. Recalling that the spin-weighted property is defined with respect to the $\hat n$ dependence, we have
\begin{align}
F_{_{-2}\gamma}(\hat q,\hat n)&\equiv \left({\slashed \partial}^2\right)^{-1}F_{\zeta}(\hat q,\hat n)\,,\nonumber\\
&=\sum_{\ell m}  (-1)^m\, A_\ell^\zeta \,_2Y_{\ell m}(\hat p)\left({\slashed \partial}^2\right)^{-1}Y_{\ell -m}(\hat n)\,,\nonumber\\
&=\sum_{\ell m}  A_\ell^\gamma \,_2Y^{\,}_{\ell m}(\hat p)\,_{-2}Y_{\ell m}^\star(\hat n)\,,
\end{align}
where
\begin{align}
\left({\slashed \partial}^2\right)^{-1}Y_{\ell m}(\hat n)=\sqrt{\frac{(\ell-2)!}{(\ell+2)!}}\,_{-2}Y_{\ell m}(\hat n)\,,
\end{align}
and
\begin{align}
A_\ell^\gamma&=\sqrt{\frac{(\ell-2)!}{(\ell+2)!}} A_\ell^\zeta =(-1)^\ell\frac{8\pi}{(\ell+1)\ell}\,,
\end{align}
\section{Calculation of the distortion response correlations}\label{app:correlations}

We now consider the correlation between distortion variables in two directions $\hat n$ and $\hat n'$, using the diagonal forms \eqref{eq:single_source_diagform}. For the case of an unpolarised SGWB, these can be obtained by integrating the conjugate product of the response functions over all GW directions \cite{allen2024pulsar} 
\begin{align}
    \Gamma^{XY}(\hat n,\hat n') &= \Re\left[\int d^2\hat q\, F^{\,}_X(\hat q,\hat n)F^\star_Y(\hat q,\hat n')\right]\,,
\end{align}
where $XY$ are combinations of distortion variables, and taking the real part is equivalent to considering an unpolarised background. This is a consequence of $F^\star(\hat q,\hat n)=F(-\hat q,-\hat n)$, i.e. conjugation of the response to a left-handed wave is equivalent to the response of a right-handed wave at the antipodal point to $\hat n$.

Below, the orthonormality property of spin-weighted spherical harmonics
\begin{align}
    \int d^2 \hat n {\,}_{s}Y^{\,}_{\ell m}(\hat n){\,}_{s}Y^\star_{\ell' m'}(\hat n) = \delta_{\ell \ell'}\delta_{m m'}\,,
\end{align}
and their addition theorem
\begin{align}
\sum_{m} {\,}_{s}Y^{\,}_{\ell m}(\hat n){\,}_{s'}Y^\star_{\ell m}(\hat n') = \frac{(2\ell+1)}{4\pi(-1)^{s+s'}} \,\mathcal{D}^\ell_{ss'}(\alpha,\beta,\xi) e^{is\xi}\,,
\end{align}
are used to simplify the expression, as done in \cite{allen2024pulsar}. The variables $\alpha$ and $\xi$ are the angles subtended by the respective local meridians at directions $\hat n$ and $\hat n'$ with the geodesic connecting the two directions and $\beta=\hat n\cdot\hat n'$. 

For the scalar (pseudo-scalar) variables $\kappa$ and $\omega$, we recover a correlation function as a sum of Legendre Polynomials $P_\ell(\beta)\equiv d^\ell_{00}(\beta)$
\begin{align}
\Gamma^{\kappa\kappa}(\hat n,\hat n')&=\Re\left[\sum_\ell \frac{2\ell+1}{4\pi}\,\left(A^\kappa_\ell\right)^2\, d_{00}^\ell\left(\beta\right)\right]\,,\nonumber\\
&=\sum_\ell \frac{2\ell+1}{4\pi}\,C_\ell^\kappa\,P_\ell(\beta)\equiv \Gamma^{\kappa\kappa}(\beta)\,,\\
\Gamma^{\kappa\omega}(\hat n,\hat n')&\equiv\Gamma^{\omega\kappa}(\hat n,\hat n')\,,\nonumber\\
&=\Re\left[\sum_\ell \mp i\frac{2\ell+1}{4\pi}\,C_\ell^\kappa\,d_{00}^\ell\left(\beta\right)\right]\,,\nonumber\\
&=0\,,
\end{align}
with $\Gamma^{\omega\omega}(\hat n,\hat n')=\Gamma^{\kappa\kappa}(\hat n,\hat n')$ and $\Gamma^{\omega\kappa}(\hat n,\hat n')=\Gamma^{\kappa\omega}(\hat n,\hat n')=0$.

The correlator between the two measurements of ${\,}_{-2}\gamma$ evaluates to
\begin{align}
\Gamma^{\gamma\gamma}(\hat n,\hat n')&=\Re\left[\sum_{\ell m}\left(A_\ell^\gamma\right)^2\,_{-2}Y_{\ell m}^\star(\hat n)\,_{-2}Y_{\ell m}(\hat n')\right]\,,\nonumber\\
&=\Re\left[\sum_{\ell}\sqrt{\frac{2\ell+1}{4\pi}}\, C_\ell^\gamma \,_{2}Y_{\ell -2}(\beta,\alpha)e^{-2i\xi}\right]\,,\nonumber\\
&=\Re\left[\sum_{\ell}\frac{2\ell+1}{4\pi}\,C_\ell^\gamma\,\mathcal{D}_{22}^\ell(\alpha,\beta,\xi)\right]\,,\nonumber\\
&=\Re\left[\sum_{\ell}\frac{2\ell+1}{4\pi}\,C_\ell^\gamma\,d_{22}^\ell(\beta) e^{-2i(\alpha+\xi)}\right]\,,\nonumber\\
&=\sum_{\ell}\frac{2\ell+1}{4\pi}\,C_\ell^\gamma\,d_{22}^\ell(\beta)\cos\left[2(\alpha+\xi)\right]\,.
\end{align}

As discussed in section~\ref{sec:correlators}, the correlation functions involving spin-s (with $s>0$) quantities are not coordinate-independent. However, the coordinate ambiguity can be avoided by defining distortion variables on locally geodesic-aligned frames with $\alpha=\xi=0$ such that
\begin{align}
    \Gamma^{\gamma\gamma}(\beta)=\sum_{\ell}\frac{2\ell+1}{4\pi}\,C_\ell^\gamma\,d_{22}^\ell(\beta)\,.
\end{align}

Similarly, the cross-correlators can be evaluated as
\begin{align}
\Gamma^{\kappa\gamma}(\hat n,\hat n')&=\Re\left[\sum_{\ell m}A_\ell^\kappa A_\ell^\gamma\,_{-2}Y_{\ell m}(\hat n)Y_{\ell m}^\star(\hat n')\right]\,,\nonumber\\
&=\Re\left[\sum_{\ell}\sqrt{\frac{2\ell+1}{4\pi}}\,
C_\ell^{\kappa\gamma}\,Y_{\ell -2}(\beta,\alpha)\right]\nonumber\\
&=\Re\left[\sum_{\ell}\frac{2\ell+1}{4\pi}\,
C_\ell^{\kappa\gamma}\,\mathcal{D}_{20}^\ell(\alpha,\beta,0)\right]\,,\nonumber\\
&=\Re\left[\sum_{\ell}\frac{2\ell+1}{4\pi}\,
C_\ell^{\kappa\gamma}\,d_{20}^\ell(\beta)\,e^{-2i\alpha}\right]\,,\nonumber\\
&=\sum_{\ell}\frac{2\ell+1}{4\pi}\,
C_\ell^{\kappa\gamma}\,d_{20}^\ell(\beta)\,\cos(2\alpha)\,,\nonumber\\
\Gamma^{\omega\gamma}(\hat n,\hat n')&=\Re\left[\sum_{\ell m}i\,A_\ell^\kappa A_\ell^\gamma\,_{-2}Y_{\ell m}(\Omega_p)Y_{\ell m}^\star(\Omega_q)\right]\,,\nonumber\\
&=\sum_{\ell}\frac{2\ell+1}{4\pi}\,C_\ell^{\kappa\gamma}\,d_{20}^\ell(\beta)\,\sin(2\alpha)\,.
\end{align}
These correlators can also be expressed in the aligned frame as
\begin{align}\label{eq:kg}
    \Gamma^{\kappa\gamma}(\beta)=\sum_{\ell}\frac{2\ell+1}{4\pi}\,
C_\ell^{\kappa\gamma}\,d_{20}^\ell(\beta)\,,
\end{align}
noting that the $d^\ell_{20}$ are related to associated Legendre functions $P^m_\ell$ as
\begin{align}
    d^\ell_{20}(\beta) = \sqrt{\frac{(\ell-2)!}{(\ell+2)!}}\, P^2_\ell(\beta)\,.
\end{align}
The $\Gamma^{\omega\gamma}$ will vanish in the aligned frame but can be verified by rotating the variables by $\pi/2$ to give the same form as \eqref{eq:kg}.

\section{Polarised backgrounds}\label{app:polarised}
The results above can be easily extended to polarised backgrounds. Consider the case where the power spectrum in \eqref{PSD_signal} is of the form
\begin{align}\mathcal{H}_{L/R}(f,\hat n)=\frac{1}{2}(1\pm p)H_{L/R}(f)\,,
\end{align}
with $0\leq p\leq 1$. The case with $p=0$ corresponds to an unpolarised background, $p=+1$ is a fully left-handed ($L$) and $p=-1$ is a right-handed ($R$) polarised background. In this case, the expectation value of the cross-correlators will be
\begin{align}
\langle \kappa(\hat n,t) \kappa^\star(\hat n',t')\rangle&=\frac{1+p}{2}\int df H_L(f)\,e^{2\pi i f(t-t')}\times\nonumber\\
&\sum_{\ell m}C_\ell^\kappa\, Y_{\ell m}^\star(\hat n)Y^{\,}_{\ell m}(\hat n')+\nonumber\\
&\frac{1-p}{2}\int df H_R(f)\,e^{2\pi i f(t-t')}\times\\&\left[\sum_{\ell m}C_\ell^\kappa \, Y_{\ell m}^\star(\hat n)Y^{\,}_{\ell m}(\hat n')\right]^\star\,,\nonumber
\end{align}
which, in the case of a fully polarised background ($p=\pm 1$) reduces to
\begin{align}
\langle \kappa(\hat n,t) \kappa^\star&(\hat n',t')\rangle_{p=\pm 1}=\int df H_{L/R}(f)\,e^{2\pi i f(t-t')}\times\nonumber\\
&\left[\sum_{\ell m}C_\ell^\kappa\,Y_{\ell m}^\star(\hat n)Y^{\,}_{\ell m}(\hat n')\right]^{(\star)}\,,\nonumber\\
&=\Gamma^{\kappa\kappa}_{L/R}(\hat n,\hat n')\int df H_{L/R}(f)\,e^{2\pi i f(t-t')}\,,\nonumber\\
\Gamma^{\kappa\kappa}_{L/R}(\hat n,\hat n')&\equiv\left[\sum_{\ell m}C_\ell^\kappa\,Y_{\ell m}^\star(\hat n)Y^{\,}_{\ell m}(\hat n')\right]^{(\star)}\,,
\end{align}
where the conjugation $(\star)$ is intended for $p=-1$ and absent for $p=+1$. Note that $\Gamma^{\kappa\kappa}_{L/R}(\hat n,\hat n')$, in principle, is complex because the polarisation tensors in the right-left-handed basis are complex quantities. However, in this specific case
\begin{align}
\Gamma^{\kappa\kappa}_{L/R}(\hat n,\hat n')=\Gamma^{\kappa\kappa}(\hat n,\hat n')\,,
\end{align}
which is defined in \eqref{eq:Gamma_kk_unpol} and is a real quantity. This means that the response function for the self-correlator of $\kappa$ coefficients is insensitive to a net polarisation, which is expected since it is not a chiral quantity.

However, considering the other correlators reveal chiral sensitivity. In particular, we find
\begin{align}
\Gamma^{\kappa\omega}_{L/R}(\hat n,\hat n')&=\sum_{\ell m}\mp \,i\, C_\ell^\kappa\,Y_{\ell m}^\star(\hat n)Y^{\,}_{\ell m}(\hat n')\,,\nonumber\\
&=\left[\Gamma^{\omega\kappa}_{L/R}(\hat n,\hat n')\right]^\star\,,
\end{align}
which now is purely imaginary and nonzero.

Considering distortion responses involving the shear variables, we first note that in \eqref{eq:vars} we calculated the response to a {\sl left-handed} polarised gravitational wave. For the spin-2 variables in that case $F_{_{+2}\gamma}(\hat q,\hat n)=0$ and $F_{_{-2}\gamma}(\hat q,\hat n)\neq 0$.
Similarly, we can show that the response functions of the ${}_{\pm 2}\gamma$ to a {\sl right-handed} polarised gravitational wave can be written in terms of the left-handed ones:
\begin{align}
F_{_{\pm 2}\gamma}^R(\hat q,\hat n)=F^\star_{_{\mp 2}\gamma}(\hat q,\hat n)\,.
\end{align}
The interpretation is that the correlation of two $_{+2}\gamma$ coefficients is solely sensitive to the right-handed part of the GW background, and the correlation of two $_{-2}\gamma$ coefficients to its left-handed part only. Cross-correlatorions between $_{+2}\gamma$ and $_{-2}\gamma$ are always zero. This implies the following expressions
\begin{align}
\langle _{\pm 2}\gamma^{\,}(\hat n,t) _{\pm 2}\gamma^\star&(\hat n',t')\rangle_{p=\mp 1}=\Gamma^{\gamma\gamma}_{R/L}(\hat n,\hat n')\times\nonumber\\
&\int df H_{R/L}(f)\,e^{2\pi i f(t-t')}\,,\nonumber\\
\langle _{\pm 2}\gamma^{\,}(\hat n,t) _{\pm 2}\gamma^\star&(\hat n',t')\rangle_{p=\pm 1}=0\,,\\
\langle _{\pm 2}\gamma^{\,}(\hat n,t) _{\mp 2}\gamma^\star&(\hat n',t')\rangle=0\,\qquad\qquad\forall\, 0\leq p\leq 1\,,\nonumber
\end{align}
with
\begin{align}
\Gamma^{\gamma\gamma}_{L/R}(\hat n,\hat n')\equiv\left[\sum_{\ell}\frac{2\ell+1}{4\pi}\,C_\ell^\gamma\,d_{22}^\ell(\beta) e^{-2i(\alpha+\xi)}\right]^{(\star)}\,.
\end{align}

Considering cross-correlations with $\kappa$ and $\omega$ we find
\begin{align}
\langle _{\pm 2}\gamma^{\,}(\hat n,t) \kappa^\star&(\hat n',t')\rangle_{p=\mp 1}=\Gamma^{\gamma\kappa}_{R/L}(\hat n,\hat n')\times\nonumber\\
&\int df H_{R/L}(f)\,e^{2\pi i f(t-t')}\,,\\
\langle _{\pm 2}\gamma^{\,}(\hat n,t) \kappa^\star&(\hat n',t')\rangle_{p=\pm 1}=0\,,\nonumber
\end{align}
with
\begin{align}
\Gamma^{\gamma\kappa}_{L/R}(\hat n,\hat n')&=\left[\sum_{\ell}\frac{2\ell+1}{4\pi}\,
C_\ell^{\kappa\gamma}\,d_{20}^\ell(\beta)\,e^{-2i\alpha}\right]^{(\star)}\,,
\end{align}
and
\begin{align}
\langle _{\pm 2}\gamma^{\,}(\hat n,t) \omega^\star&(\hat n',t')\rangle_{p=\mp 1}=\Gamma^{\gamma\omega}_{R/L}(\hat n,\hat n')\times\nonumber\\
&\int df H_{R/L}(f)\,e^{2\pi i f(t-t')}\,,\\
\langle _{\pm 2}\gamma^{\,}(\hat n,t) \omega^\star&(\hat n',t')\rangle_{p=\pm 1}=0\,,\nonumber
\end{align}
and
\begin{align}
\Gamma^{\gamma\omega}_{L/R}(\hat n,\hat n')&=\left[-i\sum_{\ell}\frac{2\ell+1}{4\pi}\,
C_\ell^{\kappa\gamma}\,d_{20}^\ell(\beta)\,e^{-2i\alpha}\right]^{(\star)}\,.
\end{align}
As one would expect, the response functions show that distortion variables can distinguish between parity-even and odd modes.

\bibliography{refs}

\end{document}